\documentclass[pra,amsmath,amsfonts,superscriptaddress,showpacs,twocolumn]{revtex4-1}
\usepackage{graphicx,graphics,times,bm,bbm,bbold,amssymb,amsmath,amsfonts,dsfont,color,xcolor,dcolumn}
\usepackage[bookmarksnumbered, colorlinks,plainpages]{hyperref}
\usepackage{graphicx}
\usepackage{dcolumn}
\usepackage{subfigure}
\usepackage{bm}
\usepackage{amsmath,amssymb}
\usepackage{bbold}
\usepackage{tikz}
\usepackage{amsmath}
\usepackage{mathtools}
\hypersetup{colorlinks,linkcolor=blue,citecolor=blue,urlcolor=blue}
\usetikzlibrary{arrows,chains,matrix,positioning,,scopes}
\usepackage{verbatim}
\graphicspath{ {./images/} }
\usepackage{verbatim}
\usepackage{float}
\usepackage{mathtools}
\usepackage{color,soul}
\usepackage{xcolor}
\usepackage{bbold}
\newtheorem{theorem}{Theorem}

\newtheorem{proposition}[theorem]{Proposition}

\newcommand{\ket}[1]{|#1\rangle}
\newcommand{\bra}[1]{\langle #1|}

\newcommand{\braket}[2]{\langle#1|#2\rangle}
\newcommand{\Tr}{{\mathrm {Tr}}}
\newcommand{\e}{{\mathrm {e}}}
\newcommand{\T}{{\mathrm {t}}}
\newcommand{\diag}{{\mathrm {diag}}}
\newcommand{\Id}{{\mathbb 1}}
\newcommand{\etal}{\textit {et al.} }

\begin{document}

\title{Mutually unbiased measurements with arbitrary purity}

\author{Mahdi Salehi}
\email{salehi.mahdi@mail.um.ac.ir}
\affiliation{Department of Physics, Ferdowsi University of Mashhad, Mashhad, Iran}
\author{Seyed Javad Akhtarshenas}
\email{akhtarshenas@um.ac.ir}
\thanks{Corresponding author}
\affiliation{Department of Physics, Ferdowsi University of Mashhad, Mashhad, Iran}
\author{Mohsen Sarbishaei}
\email{sarbishei@um.ac.ir}
\affiliation{Department of Physics, Ferdowsi University of Mashhad, Mashhad, Iran}
\author{Hakimeh Jaghouri}
\email{ha.jaghouri@alumni.um.ac.ir}
\affiliation{Department of Physics, Ferdowsi University of Mashhad, Mashhad, Iran}


\begin{abstract}
Mutually unbiased measurements are a generalization of mutually unbiased bases in which  the measurement operators need not to be rank one projectors. In a $d$-dimension space, the purity of measurement elements ranges from $1/d$ for the measurement operators corresponding to maximally mixed states to $1$ for the rank one projectors. In this contribution, we provide a class of MUM that  encompasses the  full range of purity.
Similar to the MUB in which the operators corresponding to different outcomes of the same measurement  commute mutually, our class of MUM possesses this sense of compatibility within each measurement.  The spectra of these  MUMs provides a way to  construct a class of  $d$-dimensional orthogonal matrices which leave  the vector of equal components invariant. Based on this property, and by using the MUM-based entanglement witnesses, we examine  the minimal number of measurements needed to detect entanglement of bipartite states. For a general bipartite pure state we need only two MUMs; the first one assigns  a zero mean value for all pure states, however, a complementary measurement is needed to give a negative mean value  for entangled states. Interestingly, the amount of this negative value is proportional to an entanglement monotone.

\end{abstract}



\maketitle

\section{Introduction}\label{SectionIntroduction}

Quantum measurement is one of the most puzzling aspects of quantum mechanics that makes it so counterintuitive.
This nonclassical feature arises from the fact that incompatible measurements cannot be performed simultaneously.   It makes sense therefore that the more incompatible two measurements are, the more nonclassical features they can detect. Having such a source of nonclassicality is essential in performing most protocols of the quantum information theory more efficient than the classical ones. Related to this,  mutually unbiased bases (MUBs) are a quantum ingredient that  play a fundamental role in the quantum information science. Historically, their discovery goes back to 1960, when Schwinger introduced them in connection to the notion of unitary operator bases \cite{SchwingerPNAS1960}.
Let $\{\mathcal{B}^{(0)},\mathcal{B}^{(1)},\cdots,\mathcal{B}^{(\Delta)}\}$, with $\mathcal{B}^{(b)}=\left\{\ket{e_n^{(b)}}\right\}_{n=0}^{d-1}$,  denotes a set  of $\Delta+1$ orthonormal bases on $\mathbb{C}^d$.
This set is mutually unbiased basis if and only if
\begin{equation}\label{MUB}
|\braket{e_n^{(b)}}{e_{n^\prime}^{(b^\prime)}}| = \dfrac{1}{\sqrt{d}},\quad \forall \;\; b\ne b^\prime,
\end{equation}
and $n,n^\prime\in\{0,\cdots,d-1\}$.  By words, if a system is prepared in a state belonging to one of the bases, then outcomes of measuring the system with respect to the other bases give, by no means, any information about the system's state. For a $d$-dimensional system the number of MUBs is at most $d+1$,
and if one can find a complete set of MUBs, then these bases provide a set of measurements which can be used to optimally determine the density matrix of a $d$-dimensional system \cite{WoottersAP1989,IvanovicJPA1981}. It is shown that a complete set of MUBs is an example of a rich combinatorial structure known as complex projective $2$-designs \cite{KlappeneckerPtoc2005}.
Application of MUBs is not limited to the quantum state determination  and, because of their mathematical richness, they found more application in other fields such as,   quantum error correction codes \cite{GottesmanPRA1996,CalderbankPRL1997}, quantum cryptography for secure quantum key exchange \cite{BrusPRL2002,CerfPRL2002}, quantum state tomography \cite{WoottersAP1989,AdamsonPRL2010}, and more recently  the detection of quantum entanglement \cite{SpenglerPRA2012} and the so-called "mean king's problem" \cite{EnglertPLA2001,AravindZN2003}. Therefore, constructing the complete set of MUBs is of particular importance and, despite efforts,  whether for an arbitrary $d$  there exists a complete set of MUBs  is still an open problem \cite{ButterleyPLA2007}. For particular cases where the dimension  $d$ is a prime  or a power of a prime number, it is shown that  there exists a complete set of $d+1$ mutually unbiased bases \cite{IvanovicJPA1981} and a set of two MUBs always exists \cite{DurtIJQI2010}. However, for other situations where the dimension is a composite number, the maximal number of MUBs is unknown.

By generalizing the notion of unbiasedness from bases to measurements, Kalev \etal \cite{KalevNJP2014} proposed a new notion of mutually unbiased measurements (MUMs),  which  are not necessarily rank one projectors.
Let $\{\mathcal{P}^{(0)},\mathcal{P}^{(1)},\cdots,\mathcal{P}^{({\Delta})}\}$ denotes a set of ${\Delta}+1$ positive operator-valued measures (POVMs) on $\mathbb{C}^d$, such that within each POVM $\mathcal{P}^{(b)}$,  the POVM elements are defined by  $\mathcal{P}^{(b)}=\{P_n^{(b)}\; |\; P_n^{(b)}\ge 0, \; \sum_{n=0}^{d-1}P_n^{(b)}=\Id \}$. The set  is said to be MUM if and only if \cite{KalevNJP2014} $\Tr\left[P_{n}^{(b)}\right]=1$ and
\begin{equation}\label{MUMdef1-2}
\Tr\left[P_{n}^{(b)} P_{n^{\prime}}^{(b^{\prime})}\right]=
\frac{1}{d}+\delta_{b,b^{\prime}}(\;\delta_{n,n^{\prime}}
-1/d)\left(\frac{\kappa d-1}{d-1}\right),
\end{equation}
for $n,n^\prime=0,1,\cdots,d-1$, and $b,b^\prime=0,1,\cdots,{\Delta}$. The efficiency parameter $\kappa$ is defined by $\Tr[P_{n}^{(b)} ]^2=\kappa$ and ranges from $1/d$ to $1$, just similar to the purity of quantum states on $d$-dimensional Hilbert space. Indeed, the trivial case $\kappa=1/d$ happens whenever all POVM elements are equal to the maximally mixed state $\frac{1}{d}\Id$, however, for the particular case $\kappa=1$ the POVMs reduce to the von Neumann measurements, i.e. rank one projectors on pure states, leading therefore to the particular case MUB. In view of this,  $\kappa$ can determine how close the measurement operators are to rank one projectors \cite{KalevNJP2014}.

A complete set of MUMs, with arbitrary rank, is an example of conical $2$-designs \cite{GraydonJPA2016-1} and mixed $2$-designs \cite{BransdenPRA2016}, which are both  generalizations of the complex projective $2$-designs. A set of positive semi-definite operators    $\{A_1,\cdots,A_m\}$ is a  conical $2$-design if it satisfies  the condition \cite{GraydonJPA2016-1}
\begin{eqnarray}\label{TwoConical-1}
\sum_{j=1}^{m}A_{j}\otimes A_j=k_\textrm{s}\Pi_{\textrm{sym}}+k_\textrm{a}\Pi_{\textrm{asym}},
\end{eqnarray}
where   $\Pi_{\textrm{sym}}$ and $\Pi_{\textrm{asym}}$ are projectors on the symmetric and antisymmetric subspaces of $\mathbb{C}^d\otimes\mathbb{C}^d$, respectively,  and $k_{\textrm{s}}\ge k_{\textrm{a}}\ge 0$.
In \cite{GraydonJPA2016-1},  other equivalent statements of the definition of a conical $2$-design are presented, one of the   relevant to this  work is
\begin{eqnarray}\label{TwoConical-2}
\sum_{j=1}^{m}A_{j}\otimes \overline{A}_{j}=k_{+}\Id\otimes \Id +dk_{-}\ket{\phi_{d}^{+}}\bra{\phi_{d}^{+}},
\end{eqnarray}
where  $\overline{A}$ is the conjugation of $A$, $\Id$ is the unit operator in $\mathbb{C}^d$,   $k_{\pm}=(k_{\textrm{s}}\pm k_{\textrm{a}})/2$,   and $\ket{\phi_{d}^{+}}=\dfrac{1}{\sqrt{d}}\sum_{i=0}^{d-1}\ket{ii}$ is the maximally entangled state. For a complete set of MUMs defined by  Eq.  \eqref{MUMdef1-2}, i.e.   $\Delta=d$,  the equivalent conditions \eqref{TwoConical-1} and \eqref{TwoConical-2} are   satisfied with $A_j\rightarrow P_{n}^{(b)}$, $k_{\textrm{s}}=\kappa+1$, and  $k_{\textrm{a}}=(d+1)(\kappa-1)/(d-1)$ \cite{GraydonJPA2016-1}.

Based on the  definition \eqref{MUMdef1-2}, and by defining suitable Hermitian traceless operators $F_{n}^{(b)}$,   Kalev \etal derived a complete set of $d+1$ MUMs for an arbitrary $d$-dimensional Hilbert space as \cite{KalevNJP2014}
\begin{eqnarray}\label{PnKalev}
P_{n}^{(b)}=\dfrac{1}{d}\Id+t F_{n}^{(b)},
\end{eqnarray}
where $t$ is a free parameter, its range
\begin{eqnarray}\label{tRanges}
\dfrac{-1}{d} \dfrac{1}{\lambda_{\max}} \leq t \leq \dfrac{1}{d} \dfrac{1}{\vert \lambda_{\min} \vert},
\end{eqnarray}
guarantees the positivity  of  $P_{n}^{(b)}$.
Here, $\lambda_{\min}=\min_{b} \lambda_{\min}^{b}$ and  $\lambda_{\max}=\max_{b} \lambda_{\max}^{b}$, where $\lambda_{\min}^{b}$ and $\lambda_{\max}^{b}$ are respectively the smallest and the largest eigenvalue of the operators $F_{n}^{(b)}$ with $n=0,1,\cdots,d-1$.
The efficiency parameter $\kappa $ is simply related to $t$ as $\kappa=1/d + t^{2}(1+\sqrt{d})^2(d-1)$, and due to restriction \eqref{tRanges}, $\kappa$ is also restricted. This, in turn, implies that in order to have MUB, i.e. $\kappa=1$, parameter $t$ should satisfy $t^2=1/[d(1+\sqrt{d})^2]$, meaning that we have to find  a set of operators $F_n^{(b)}$ such that $\min\{\lambda_{\max},\vert \lambda_{\min} \vert\}=1+1/\sqrt{d}$. Whether there exists such a set of operators is of particular importance and is closely related to the  question that whether a complete set of MUB exists.
The $d(d+1)$ operators $F_{n}^{(b)}$ are defined as a suitable linear combination of $d^2-1$ Hermitian, traceless orthonormal operator basis of the space of Hermitian, traceless operators. The  form of linear combination, as presented in \cite{KalevNJP2014},  guarantees that relation \eqref{MUMdef1-2} is satisfied. However, as it is shown in \cite{KalevNJP2014}, choosing generalized Gell-Mann operator basis gives the optimal value for efficiency parameter as $\kappa_{\textrm{opt}}=1/d+2/d^2$, which is far from its upper bound $\kappa=1$, unless for $d=2$.
This is not surprising as commutativity of elements of  each POVM $\mathcal{P}^{(b)}$ is an essential feature of being MUB ($\kappa=1$), but the form provided in Eq. \eqref{PnKalev} lacks this feature except for the particular case  $d=2$. This comes from the fact that for a given POVM $b$, the set of $d$ operators $F_n^{(b)}$  do not commute mutually, i.e.  $[F_n^{(b)},F_{n^\prime}^{(b)}]\ne 0$ for $n\ne n^\prime$ with $n,n^\prime=0,1,\cdots,d-1$. To overcome this limitation, one can look for a set of operators $F_n^{(b)}$ that (i) either  they depend on the efficiency  parameter $\kappa$ such that the commutators vanish for $\kappa=1$,  (ii) or we look for a set of operators that depend on $\kappa$, however their commutators vanish for all allowed values of $\kappa$.

In this paper, we follow the second approach and provide a class of MUM for all values of $\kappa\in [1/d,1]$. For this purpose,  we   require the following  conditions to be satisfied for our class,  both inherited from MUB ($\kappa=1$);  (i) within each POVM, the POVM elements commute mutually and (ii) the spectra of all POVM elements is the same, regardless of which POVM the elements belong to. Obviously, within  these conditions the problem of finding the complete set of MUM for an arbitrary $d$ becomes equivalent to the currently open problem of finding the complete set of MUB. This is, however,  the price that we have to pay in order to achieve a family of MUMs with purity ranging over all allowed values of the efficiency parameter $\kappa$.
 The spectra of POVMs can be used to construct a particular class of $d$-dimensional orthogonal matrices. These orthogonal matrices possess the property of leaving a vector of equal components invariant.   Based on the notion of  MUM-based entanglement witnesses, we invoke such a class of MUM, to detect entanglement of bipartite states. We show that to detect entanglement of a general bipartite pure state, we need only two MUMs; The first one assigns  a zero mean value for all pure states, however, a complementary measurement is needed to give a negative mean value  for entangled states. Interestingly, the amount of this negative value is proportional to an entanglement monotone.  For bipartite mixed states, however,   the  number of MUMs which is needed to detect entanglement increases in general. We examine this by providing some examples and show that although for a full rank state, such as isotropic state, a complete set of MUMs is needed to completely detect entanglement, for mixed states with low-dimensional support a smaller set of MUMs suffices to capture entanglement.
 The remaining of the paper is structured as follows. In section \ref{SectionMUMs}, we construct  our class of MUMs with arbitrary purity. Section \ref{SectionWitness} is devoted to use this class of MUMs to detect entanglement of bipartite states. The paper is concluded in section \ref{SectionConclusion} with a brief conclusion.

\section{A class of MUMs with arbitrary purity}\label{SectionMUMs}
Consider  a set of ${\Delta}+1$  POVMs  $\mathcal{P}^{(b)}$,  $b\in \{0,1,\cdots,{\Delta}\}$, such that for each $b$ the  POVM  elements are    described by
\begin{eqnarray}\label{MUMdef2}
P_{n}^{(b)}=\frac{1}{d}\Id+M_{n}^{(b)},
\end{eqnarray}
for  $n=0,\cdots,d-1$. Above,  $M_{n}^{(b)}$ is a Hermitian traceless matrix defined by $M_{n}^{(b)}=\boldsymbol{r}_{n}^{(b)} \cdot \boldsymbol{\lambda}$,  where $\boldsymbol{\lambda}=(\lambda_1,\cdots,\lambda_{d^2-1})^\T$ denotes a vector constructed by  generalized  Gell-Mann  basis, and  $\boldsymbol{r}^{(b)}_n=(r_{n,1}^{(b)},\cdots,r_{n,d^2-1}^{(b)})^\T$  is the Bloch vector corresponding to the POVM $P_{n}^{(b)}$.
The MUM condition \eqref{MUMdef1-2} requires the following condition on the traceless matrices $M_{n}^{(b)}$
\begin{equation}\label{TrMMp-1}
\Tr\left[M_{n}^{(b)} M_{n^{\prime}}^{(b^{\prime})}\right]= \delta_{b,b^{\prime}}(\;\delta_{n,n^{\prime}}
-1/d)\left(\frac{\kappa d-1}{d-1}\right).
\end{equation}
Regarding the definition \eqref{MUMdef1-2}, Eq. \eqref{MUMdef2} provides the most general form for MUM, however, in the following we restrict ourselves to the case that; (i) within each   POVM $\mathcal{P}^{(b)}$, the POVM elements commute mutually, i.e. $[M_n^{(b)},M_{n^\prime}^{(b)}]=0$ for all $n,n^\prime=0,\cdots,d-1$,   and (ii) the spectra of all POVM elements are  the same, regardless of which POVM the elements belongs to.  This includes, in particular,  the special case where the POVM is  a projective measurement.

Suppose the computational  basis of $\mathbb{C}^d$, $\{\ket{e_i}\}_{i=0}^{d-1}$,  is chosen in such a way that all elements of a particular POVM, say
$\mathcal{P}^{(0)}$,  are diagonal, i.e.
\begin{eqnarray}\label{mum}
P_{n}^{(0)}=\frac{1}{d}\Id+M_{n}^{(0)},
\end{eqnarray}
where $M_{n}^{(0)}=\diag\{([\mu_{n}]_{0},\cdots,[\mu_{n}]_{d-1}\}$, for $n=0,\cdots,d-1$,  is a diagonal matrix corresponding to the $n$th element of $\mathcal{P}^{(0)}$. Accordingly, conditions $\Tr{[P_{n}^{(0)}]}=1$ and $\Tr{[P_{n}^{(0)}]^2}=\kappa$ read
\begin{equation} \label{POVMconditions1}
\qquad\sum_{i=0}^{d-1} [\mu_{n}]_{i}=0, \qquad \sum_{i=0}^{d-1} [\mu_{n}]_{i}^{2}=\kappa - \frac{1}{d},
\end{equation}
respectively, for $n=0,\cdots,d-1$. Adding to these, the positivity $0\leqslant P_{n}^{(0)}\leqslant \Id$, for $n=0,\cdots,d-1$,  and the completeness $\sum_{n=0}^{d-1}P_{n}^{(0)}=\Id$ conditions, we get
\begin{equation}\label{POVMconditions2}
-\frac{1}{d} \leqslant [\mu_{n}]_{i}\leqslant \frac{d-1}{d}, \qquad \sum_{n=0}^{d-1}[\mu_{n}]_{i}=0,
\end{equation}
for $i=0,\cdots,d-1$, respectively.

A comparison of the first condition of Eq. \eqref{POVMconditions1} with the second condition of Eq. \eqref{POVMconditions2} shows that one can choose the  diagonal matrices  $M_{n}^{(0)}$ ($n=0,1,\cdots,d-1$) such that they are mutually equivalent up to the location of their  diagonal elements. By defining  the transposition operator $\mathcal{S}_n$ with entries $[\mathcal{S}_n]_{ij}=\braket{i\oplus n}{j}$, where  $\oplus$ means addition modulo $d$,  one can write $M_{n}^{(0)}=\mathcal{S}_n M_{0}^{(0)}\mathcal{S}_n^{\T}$ for $n=0,\cdots,d-1$. In view of this, we can write $[M_n^{(0)}]_{jj}=\mu_{n\oplus j}$, or equivalently, $M_n^{(0)}=\diag\{\mu_{n},\mu_{n\oplus 1},\mu_{n\oplus 2}\cdots,\mu_{n\oplus d-1}\}$. With this convention, it is clear that $[M_n^{(0)}]_{jj}=[M_j^{(0)}]_{nn}=\mu_{n\oplus j}$. The eigenvalues  satisfy (see Eqs. \eqref{TrMMp-1}, \eqref{POVMconditions1} and \eqref{POVMconditions2})
\begin{equation} \label{POVMconditions3}
\sum_{j=0}^{d-1} \mu_{n\oplus j}=\sum_{n=0}^{d-1} \mu_{n\oplus j}=0, \quad -\frac{1}{d} \leqslant \mu_{n\oplus j}\leqslant \frac{d-1}{d},
\end{equation}
and
\begin{equation} \label{POVMconditions4}
\sum_{j=0}^{d-1} \mu_{n\oplus j}\mu_{n^\prime\oplus j}=(\;\delta_{n,n^{\prime}}
-1/d)\left(\frac{\kappa d-1}{d-1}\right).
\end{equation}
It follows that not all eigenvalues $\mu_{n\oplus j}$ are independent. In Appendix \ref{App1}, we show that  the number of independent eigenvalues reduces  to  $N=d-\left\lfloor d/2\right\rfloor=\left\lfloor\frac{d-1}{2}\right\rfloor$, where $\left\lfloor x/2\right\rfloor$ denotes the integral part of $x/2$, i.e.  $\left\lfloor x/2\right\rfloor=x/2$ if $x$ is even and  $\left\lfloor x/2\right\rfloor=(x-1)/2$ if $x$ is odd.

 It is worth mentioning that the set of eigenvalues of $M_n^{(0)}$ can be used to construct a one-parameter family of orthogonal matrices. To see this, define $q=\sqrt{\frac{1-1/d}{\kappa-1/d}}$ (for $1/d<\kappa\le 1$) and note that two real $d$-component vectors $v_0=\left(q\mu_{0}+1/d,q\mu_{1}+1/d,\cdots,q\mu_{d-1}+1/d\right)^{\T}$  and $v_1=\left(q\mu_{d-1}+1/d,q\mu_{0}+1/d,\cdots,q\mu_{d-2}+1/d\right)^{\T}$ are  normalized and orthogonal. The following result is then  obtained by extending this to all  vectors $v_j$ ($j=0,1,\cdots,d-1$) with components $\{q\mu_{n\oplus d-j}+1/d\}_{n=0}^{d-1}$ .
 \begin{proposition}\label{PropQ}
 The $d\times d$ matrix $\mathcal{Q}$ defined by
\begin{eqnarray}\label{Qmatrix}
\mathcal{Q}_{nj}&=&q[M_n^{(0)}]_{d-j,d-j}+1/d
\\ \nonumber
&=&q\mu_{n\oplus d-j}+1/d,
\end{eqnarray}
 is orthogonal, i.e. $\mathcal{Q}\mathcal{Q}^\T=\Id_d$. Moreover, it satisfy  $\sum_{n=0}^{d-1}\mathcal{Q}_{nj}=\sum_{j=0}^{d-1}\mathcal{Q}_{nj}=1$  implies  that $\mathcal{Q}$ leaves the vector $\boldsymbol{n}_\ast=(1,1,\cdots,1)/\sqrt{d}$ invariant, i.e.  $\mathcal{Q}\boldsymbol{n}_\ast=\boldsymbol{n}_\ast$.
\end{proposition}
As an explicit example let us  consider the  case $d=3$.
In this case,  we have
\begin{eqnarray}
\mathcal{Q}=\left(\begin{array}{ccccc} q\mu_{0}+1/3 && q\mu_{2}+1/3 && q\mu_{1}+1/3 \\
q\mu_{1}+1/3 && q\mu_{0}+1/3 && q\mu_{2}+1/3 \\
q\mu_{2}+1/3 && q\mu_{1}+1/3 && q\mu_{0}+1/3
 \end{array}\right).
\end{eqnarray}
Here  $q=\sqrt{2/(3\kappa-1)}$ and, as we mentioned above, we have only one independent eigenvalue, say $\mu_{0}$. The  other two eigenvalues can be expressed as  $\mu_{1,2}=[-\mu_0\pm \sqrt{2\kappa-2/3-3\mu_0^2}]/2$.  In this simple case,  the eigenvalues can be expressed in terms of single independent parameter $\phi$, i.e. for a given purity $\kappa$, one can write $\mu_0=\sqrt{2/3}\sqrt{\kappa-1/3}\cos{\phi}$  and  $\mu_{1,2}=\sqrt{2/3}\sqrt{\kappa-1/3}\cos{(\phi\pm 2\pi/3)}$, where $0\le \phi \le 2\pi$.

Because of high dependency between eigenvalues, it is not always possible to express the set of eigenvalues as  explicit functions of  independent parameters. Equations \eqref{POVMconditions3} and \eqref{EigenCondition1}, however,   allow us to  write two of them in terms of the sum and the sum of square of the others
\begin{eqnarray}\label{Fpm}
\mu_{p,m}=-\frac{1}{2}S\pm\frac{1}{2}\sqrt{2(\kappa-1/d)-2T-S^2},
\end{eqnarray}
where we have defined $S=\sum_{l\ne p,m}^{d-1}\mu_l$ and $T=\sum_{l\ne p,m}^{d-1}\mu^2_l$. Recall that $1/d\le \kappa\le 1$, the lower bound $\kappa=1/d$ happens  when  $\mu_n(\kappa=1/d)=0$ for all $n=0,\cdots,d-1$. The upper bound $\kappa=1$ occurs when  $\mu_n(\kappa=1)=-1/d$ for all $n$ except $n=p$.
  Note that although the negative region $[-1/d,0]$ is allowed for all  $\mu_l$s at the same time, the positive region is not  accessible to all eigenvalues simultaneously. For example, if one of the eigenvalues reaches the maximum value $1-1/d$, all others must take their minimum value $-1/d$.

{\it Constructing a set of MUMs.---}So far, we have provided a form for the POVM $\mathcal{P}^{(0)}$. In order to construct a set of MUMs (or MUBs), we have to find POVMs $\mathcal{P}^{(b)}$ for $b=1\cdots,{\Delta}$ where  ${\Delta}\le d$.
As our set of MUMs possesses the same spectra, we are looking  therefore  for some unitaries  $U^{(b,b^\prime)}$, taking $\mathcal{P}^{(b^\prime)}$ and generate $\mathcal{P}^{(b)}$, i.e. $P_n^{(b)}=U^{(b,b^\prime)}P_n^{(b^\prime)}{U^{(b,b^\prime)}}^\dagger$, for all $n=0,\cdots,d-1$ and $b,b^\prime=0,\cdots,{\Delta}$. Clearly, $U^{(b,b)}=\Id$.
Defining $\ket{e_l^{(b)}}=U^{(b,b^\prime)}\ket{e_l^{(b^\prime)}}$ for $l=0\cdots,d-1$, and noting that  $\ket{e_l^{(0)}}=\ket{e_l}$ is our computational basis, one can write the spectral decomposition of $M_{n}^{(b)}$ as
\begin{equation}\label{Mnb}
M_{n}^{(b)}=\sum_{l=0}^{d-1}\mu_{n\oplus l}\ket{e_l^{(b)}}\bra{e_l^{(b)}}.
\end{equation}
As we mentioned above, in  the computational  basis $\{\ket{e_i}\}_{i=0}^{d-1}$,  matrices $M_n^{(0)}$ (for $n=0,\cdots,d-1$) are diagonal. They can, therefore,  expressed in terms of the Cartan  subalgebra  of the group of unitary matrices $SU(d)$, i.e. the maximal abelian subalgebra generated by $d-1$ Cartan generators
\begin{eqnarray}\nonumber
\lambda_{j}=\frac{1}{\sqrt{2j(j+1)}}\left[\sum_{m=1}^{j}\ket{e_m}\bra{e_m}-j\ket{e_{j+1}}\bra{e_{j+1}}\right],
\end{eqnarray}
for $j=1,\cdots,d-1$. Central to our  discussion is that Cartan subalgebra is not unique, meaning that  any conjugate of Cartan subalgebra by an arbitrary element of the group is another Cartan subalgebra, i.e. $g\mathfrak{h} g^{-1}=\mathfrak{h}$ for  $g\in G$. In view of this, one can define conjugates of Cartan subalgebra as $\mathfrak{h}^{(b)}=U^{(b)}\mathfrak{h}^{(0)} {U^{(b)}}^\dagger$, where we have defined  $U^{(b,0)}=U^{(b)}$ for the sake of simplicity, and $\mathfrak{h}^{(0)}$ denotes the Cartan subalgebra which is diagonal in the computational basis.

In the light of  this, we are looking for some conjugates of Cartan subalgebra related by an automorphism of the algebra, in such a way that the conjugates are mutually orthogonal in the sense of vector space inner product.  More precisely, let $\mathcal{V}$ denotes the underlying space of the Lie algebra $\mathfrak{su}(d)$, and suppose $\mathcal{V}^{(0)}\subset \mathcal{V}$ to be the subspace corresponding to the Cartan subalgebra $\mathfrak{h}^{(0)}$.
Two questions arise now. First, how can we find subspaces $\mathcal{V}^{(b)}=U^{(b)} \mathcal{V}^{(0)} {U^{(b)}}^\dagger$, by means of unitary transformations $U^{(b)}$,   such that  $\mathcal{V}^{(b)}\perp \mathcal{V}^{(0)}$ and   for any $b\ne b^\prime$ we have $\mathcal{V}^{(b)}\perp \mathcal{V}^{(b^\prime)}$. Here orthogonality between two subspaces means that for any  $X\in \mathcal{V}^{(b)}$ and $Y\in \mathcal{V}^{(b^\prime)}$,  we have $\Tr(X^\dagger Y)=0$ for $b\ne b^\prime$.  Second, is it possible to write the Lie algebra vector space $\mathcal{V}$ as an orthogonal direct sum of isomorphic subspaces $\mathcal{V}^{(b)}=U^{(b)} \mathcal{V}^{(0)} {U^{(b)}}^\dagger$, i.e.
\begin{equation}
\mathcal{V}=\bigoplus_{b=0}^{d} \mathcal{V}^{(b)},
\end{equation}
with $U^{(0)}=\Id$.  Such a decomposition, if exists, can be used  to find a complete set of MUBs for a $d$-dimensional quantum system.

Now let us see the form of these unitary transformations. In the light of Eq. \eqref{Mnb}, Eq. \eqref{TrMMp-1} reads
\begin{equation}\label{TrMMp-2}
\sum_{j=0}^{d-1}\sum_{j^\prime=0}^{d-1}\mu_{n\oplus j}[B^{(b,b^\prime)}]_{jj^\prime}\mu_{n\oplus j^{\prime}}=0, \quad \textrm{for} \quad b\ne b^\prime,
\end{equation}
and   $n=0,\cdots,d-1$. Here,  $B^{(b,b^\prime)}$ is a unistochastic matrix with entries  $[B^{(b,b^\prime)}]_{jj^\prime}=|[U^{(b,b^\prime)}]_{jj^\prime}|^2$ where  $U^{(b,b^\prime)}=\sum_{j=0}^{d-1}\ket{e_j^{(b)}}\bra{e_j^{(b^\prime)}}$ is a unitary matrix with entries $[U^{(b,b^\prime)}]_{jj^\prime }=\bra{e_{j}^{(b^\prime)}}U^{(b,b^\prime)}\ket{e_{j^\prime}^{(b^\prime)}}=\braket{e_{j}^{(b^\prime)}}{e_{j^\prime}^{(b)}}$.
Since the mutually unbiased condition requires  $|\braket{e_{j}^{(b)}}{e_{j^\prime}^{(b^\prime)}}|=1/{\sqrt{d}}$, we therefore arrive at the following condition for MUB;
A necessary and sufficient condition for a set of ${\Delta}+1$  orthonormal basis $\{\ket{e_j^{(b)}}\}_{j=0}^{d-1}$, $b=0,\cdots,{\Delta}$, to form a  mutually unbiased bases is that  there exist unistochastic matrices $B^{(b,b^\prime)}$ with entries $[B^{(b,b^\prime)}]_{jj^\prime}=1/d$, for $0 \le b^\prime < b \le {\Delta}$.

{\it Bloch vector representation.---}Let us now turn our attention on the Bloch vectors of  $\mathcal{P}^{(b)}$.  By  inserting $M_{n}^{(b)}=\boldsymbol{r}_{n}^{(b)} \cdot \boldsymbol{\lambda}$ in  Eq. \eqref{TrMMp-1} we get
\begin{equation}\label{BV-InnerP}
\boldsymbol{r}^{(b)}_n\cdot \boldsymbol{r}^{(b^\prime)}_{n^\prime}= 2\delta_{b,b^{\prime}}(\kappa -1/d)\left[\frac{d\;\delta_{n,n^{\prime}}
-1}{d-1}\right],
\end{equation}
where we have used $\Tr\{\lambda_k\lambda_l\}=\delta_{kl}/2$. This  equation is equivalent to Eq. \eqref{TrMMp-1} and implies that  all Bloch vectors have the same length $|\boldsymbol{r}^{(b)}_n|=\sqrt{2}\sqrt{\kappa-1/d}$, and that    within each POVM $\mathcal{P}^{(b)}$ the  angle between two Bloch vectors  $\boldsymbol{r}^{(b)}_n$ and $\boldsymbol{r}^{(b)}_{n^\prime}$ is the same for all pairs $n\ne n^\prime$,  i.e.  $\theta^{(b)}_{n,n^\prime}=\arccos\{\frac{-1}{d-1}\}$. Therefore, associated to each POVM $\mathcal{P}^{(b)}$, the  Bloch vectors $\{\boldsymbol{r}^{(b)}_n\}_{n=0}^{d-1}$    form a regular $(d-1)$-simplex $\boldsymbol{\Delta}^{(b)}_{d-1}$ in $\mathbb{R}^{d^2-1}$. Moreover, the mutually unbiasedness condition requires that the simplexes corresponding to two different POVMs be orthogonal, i.e. $\boldsymbol{\Delta}^{(b)}_{d-1}\perp \boldsymbol{\Delta}^{(b^\prime)}_{d-1}$ for $b\ne b^\prime$.  Such simplexes live effectively in the $(d-1)$-dimensional subspaces of $\mathbb{R}^{d^2-1}$.
In particular, for the simplex associated with the POVM $\mathcal{P}^{(0)}$, the diagonal nature  of $\{M_{n}^{(0)}\}$ for ($n=0,\cdots,d-1$)  ensures that  the corresponding  Bloch vector $\boldsymbol{r}^{(0)}_n=(r_{n,1}^{(0)},\cdots,r_{n,d^2-1}^{(0)})^\T$  has the nonzero components along the  Cartan subalgebra only. These nonzero components are  given by  $\sqrt{2/(j(j+1))}\left[\sum_{m=1}^j\mu_{n\oplus m}-j\mu_{n\oplus j+1}\right]$ for $j=1,\cdots,d-1$.

 Clearly, all other simplexes can be obtained by rotating the original simplex. In general, starting from the Bloch vectors $\{\boldsymbol{r}^{(b^\prime)}_n\}_{n=0}^{d-1}$, one can always find the rotation matrix $R^{(b,b^\prime)}$ to generate $\{\boldsymbol{r}^{(b)}_n\}_{n=0}^{d-1}$ such that $\boldsymbol{\Delta}^{(b^\prime)}_{d-1}\perp \boldsymbol{\Delta}^{(b)}_{d-1}$, i.e.  $[\boldsymbol{r}^{(b)}_n]_k=\sum_{l=1}^{d^2-1}R^{(b,b^\prime)}_{kl}[\boldsymbol{r}^{(b^\prime)}_n]_l$. We demand, however, that the matrix elements of  $R^{(b,b^\prime)}$ comes from unitaries $U^{(b,b^\prime)}$ via \cite{ZyczkowskiBook2017}
 \begin{equation}\label{RU}
 R^{(b,b^\prime)}_{kl}=\Tr\{\lambda_k U^{(b,b^\prime)}\lambda_l {U^{(b,b^\prime)}}^\dagger\},
 \end{equation}
where $U^{(b,b^\prime)}$ are required to satisfy the conditions given by \eqref{TrMMp-2}. The rotations defined by Eq. \eqref{RU} form a subgroup of $SO(d^2-1)$. Although it is always possible to find some rotations $SO(d^2-1)$ that generate a complete set of $d+1$ mutually orthogonal simplexes from the original simplex, whether for an arbitrary $d$ there exists a set of rotations expressed by Eq. \eqref{RU} is still an open problem.

For further use,  we consider the case $d=3$, for which  the Gell Mann matrices $\{\lambda_i\}_{i=1}^{8}$ is a basis for $\mathfrak{su}(3)$, and the corresponding two-dimensional Cartan subalgebra $\mathfrak{h}^{(0)}$  is generated  by $\lambda_3=\frac{1}{\sqrt{3}}\diag\{1,-1,0\}$ and $\lambda_8=\frac{1}{\sqrt{6}}\diag\{1,1,-2\}$.
In this case,  three  orthogonal conjugations of Cartan subalgebra  are generated by $\{\lambda_3^{(b)}, \lambda_8^{(b)}\}$ for $b=1,2,3$, such that
$\lambda_i^{(b)}=U^{(b)}\lambda_i{U^{(b)}}^\dagger$ for $i=3,8$, where $U^{(1)}$ is nothing but the Fourier transformation
\begin{eqnarray}\label{U1-3}
U^{(1)}=\frac{1}{\sqrt{3}}
\begin{pmatrix}
1&1&1\\1&\omega &\omega^{2}\\1&\omega^{2}&\omega
\end{pmatrix},
\end{eqnarray}
and $U^{(2)}=VU^{(1)}$, and   $U^{(3)}=VU^{(2)}$. Here   $V=\diag\{1,\omega,\omega\}$ and  $\omega=\e^{2\pi i/3}$.

\section{Witnessing entanglement using MUM}\label{SectionWitness}
Among the various applications of MUBs, their ability in detecting quantum entanglement is one of the most recent and interesting ones.
In a bipartite system, correlations could improve the predictability on the outcomes of a measurement on one side  when we know the outcome of a measurement on the other side.  Following this notion and given any  set of ${\Delta}+1$ mutually unbiased bases     $\mathcal{B}^{(b)}=\{\ket{e_n^{(b)}}\}$ and $\mathcal{B}^{\prime {(b)}}=\{\ket{e^{\prime (b)}_n}\}$ ($b=0,1,\cdots,{\Delta}$, see Eq. \eqref{MUB}) for parts $A$ and $B$, respectively, the authors of \cite{SpenglerPRA2012} have shown that for all separable states the following condition holds
\begin{equation}\label{Wit1}
I_{{\Delta}+1}=\sum_{b=0}^{{\Delta}}\sum_{n=0}^{d-1}\Tr[E_n^{(b)}\otimes E^{\prime (b)}_{n} \rho ]\le 1+\frac{{\Delta}}{d},
\end{equation}
where $E_n^{(b)}=\ket{e_n^{(b)}}\bra{e_n^{(b)}}$ and $E^{\prime (b)}_{n}=\ket{e^{\prime (b)}_n}\bra{e^{\prime (b)}_n}$ are projectors on the orthonormal bases  $\mathcal{B}^{(b)}$ and $\mathcal{B}^{\prime {(b)}}$, respectively.  It turns out  that violation of inequality  \eqref{Wit1} is a criterion for entanglement detection. Remarkably, the efficiency of the above criterion  depends on the number ${\Delta}+1$ of MUBs. Actually, it has been  shown that to verify entanglement of pure states  two MUBs suffice, however,  one needs more MUBs  to detect entanglement of noisy states, i.e.  the noise robustness of the criterion  increases with the number ${\Delta}+1$ of MUBs \cite{SpenglerPRA2012}.

Instead of using a set of ${\Delta}+1$ MUBs, the authors of \cite{ChenPRA2014} have followed  the notion of  a complete set of $d+1$ MUMs and have shown that for all separable states  the following condition holds
\begin{equation}\label{Wit2}
J(\rho) = \sum_{b=0}^{d}\sum_{n=0}^{d-1}\Tr[P_{n}^{(b)}\otimes P_{n}^{\prime(b)} \rho]\le 1+\kappa.
\end{equation}
Above $P_{n}^{(b)}$ and $P_{n}^{\prime(b)}$ are any two complete sets of MUMs with the same efficiency parameter $\kappa$ of parts $A$ and $B$, respectively,  and are  defined previously   by Eq. \eqref{PnKalev}.

The most general tool to detect entanglement is the so-called entanglement witness which is related closely to the more fundamental concept of the positive maps, i.e., the linear  map $\Phi:\mathcal{B}(\mathcal{H}_1)\rightarrow \mathcal{B}(\mathcal{H}_2)$ such that $\Phi X\ge 0\; \forall X\ge 0$, where $\mathcal{B}(\mathcal{H})$ denotes the space of bounded operators acting on the Hilbert space  $\mathcal{H}$. For a given positive but not completely positive  map $\Phi$, one can construct the corresponding entanglement witness $W=(d-1)(\Id\otimes \Phi)\ket{\phi_d^{+}}\bra{\phi_d^{+}}$, where $\ket{\phi_d^{+}}=\frac{1}{\sqrt{d}}\sum_{i=0}^{d}\ket{ii}$. It turns out that $W$ has nonnegative expectation value on any separable state $\rho$, i.e. $\Tr{\rho W}\ge 0$  for all separable states. Accordingly, a state for which $\Tr{\rho W}<0$ is entangled.
Following this notion and for any given set of MUBs,  the authors of \cite{ChruscinskiPRA2018} have recently introduced the following  positive and trace preserving  map
\begin{equation}
\phi X = \phi_{*}X - \dfrac{1}{d-1}\sum_{b=0}^{{\Delta}}\sum_{k,l=0}^{d-1}\mathcal{O}_{k,l}^{(b)}\Tr(\tilde{X}E_{l}^{(b)})E_{k}^{(b)},
\end{equation}
where $\phi_{*}X=\dfrac{1}{d}\Id \Tr X$ defines the completely depolarizing channel and  $\tilde{X}=X-\phi_{*}X$ denotes the traceless part of X. Also, $\mathcal{O}^{(b)}$ is a set of orthogonal rotation in $\mathbb{R}^d$ around the axis $\boldsymbol{n}_\ast=(1,1,\cdots,1)/\sqrt{d}$, so that $\mathcal{O}^{(b)}\boldsymbol{n}_\ast=\boldsymbol{n}_\ast$.
The corresponding entanglement witness reads as \cite{ChruscinskiPRA2018}
\begin{eqnarray}\label{WitnessE}
W= \dfrac{d +{\Delta}}{d} \Id \otimes \Id - \sum_{b=0}^{{\Delta}}\sum_{k,l=0}^{d-1}\mathcal{O}_{kl}^{(b)} \overline{E}_{l}^{(b)}\otimes E_{k}^{(b)},
\end{eqnarray}
where $\overline{E}_{l}^{(b)}$ is the conjugation of $E_{l}^{(b)}$, and ${\Delta}\in \{0,1,\cdots,d\}$. A generalization of the above result in terms of MUMs is given in \cite{LiITP2019}
\begin{eqnarray}\label{WitnessP}
W(\kappa)= \dfrac{d\kappa +{\Delta}}{d} \Id \otimes \Id - \sum_{b=0}^{{\Delta}}\sum_{k,l=0}^{d-1}\mathcal{O}_{kl}^{(b)} \overline{P}_{l}^{(b)}\otimes P_{k}^{(b)}.
\end{eqnarray}
Note that for a complete set of MUMs, i.e. $\Delta=d$, the simplest case corresponding to $\mathcal{O}_{kl}^{(b)}=\delta_{kl}$ enables one to apply  Eq. \eqref{TwoConical-2} to find the well known witness $W=(1/d)\Id\otimes\Id-\ket{\phi_{d}^{+}}\bra{\phi_{d}^{+}}$ (up to an  overall constant factor).   This witness  detects  entanglement of the state $\ket{\phi_{d}^{+}}$ and of the noisy states that are close to $\ket{\phi_{d}^{+}}$.

Now, to proceed with the  witness  \eqref{WitnessP}, it is useful to use Eq. \eqref{MUMdef2} and  provide a simplified form for the expectation value of the witness over a general bipartite state.
\begin{proposition}
For any $d\times d$ bipartite state $\rho$, the expectation value of the witness given above can be written as
\begin{equation}\label{Prop1}
\Tr\{\rho W(\kappa)\}=(\kappa -1/d)-\Tr\{\rho \mathcal{M}(\kappa)\},
\end{equation}
where $\mathcal{M}(\kappa)$ is a traceless matrix given by
\begin{equation}\label{Mkappa}
\mathcal{M}(\kappa)=\sum_{b=0}^{{\Delta}}\sum_{k,l=0}^{d-1}\mathcal{O}_{kl}^{(b)} \overline{M}_{l}^{(b)}\otimes M_{k}^{(b)}.
\end{equation}
\end{proposition}
The proof is easy by virtue of $\sum_{k=0}^{d-1}\mathcal{O}_{kl}^{(b)}=\sum_{l=0}^{d-1}\mathcal{O}_{kl}^{(b)}=1$ and $\sum_{k=0}^{d-1}M_{k}^{(b)}=0$.
According to Eq. \eqref{Prop1},  $W(\kappa)$ detects entanglement of $\rho$ if and only if $\Tr\{\rho \mathcal{M}(\kappa)\}>\kappa-1/d$,  otherwise $\rho$  is not entangled or its entanglement cannot be detected by $W(\kappa)$. As we expect, for the trivial case $\kappa=1/d$ we have  $\mathcal{M}(1/d)=0$, so $W(\kappa)$ cannot detect  any entanglement, or equivalently, it is not entanglement witness at all. An interesting property of this condition is that although $\mathcal{M}(\kappa)$ depends on both the purity $\kappa$ of the POVMs and the number of POVMs in the  set of  MUM, the lower bound is independent of the number of POVMs.

 The interpretation of Eq. \eqref{Prop1} becomes more  clear if one uses  $M_n^{(b)}=U^{(b)}M_n^{(0)}{U^{(b)}}^\dagger$ for some unitary matrices $U^{(b)}$ satisfying Eq. \eqref{TrMMp-2},  and  writes down
\begin{equation}\label{RhoMkappa2}
\Tr\{\rho \mathcal{M}(\kappa)\}=\sum_{b=0}^{{\Delta}}\Tr\{ \rho^{(b)}\mathcal{M}^{(b)}(\kappa)\},
\end{equation}
where $\rho^{(b)}=({\overline{U}^{(b)}}^\dagger\otimes {U^{(b)}}^\dagger) \rho(\overline{U}^{(b)}\otimes U^{(b)})$ and $\mathcal{M}^{(b)}(\kappa)=\sum_{k,l=0}^{d-1}\mathcal{O}_{kl}^{(b)}\overline{M}_{l}^{(0)}\otimes M_{k}^{(0)}$. In the light of this, one can see that to detect entanglement of the state $\rho$, we have to find a set of $\Delta+1$ mutually locally unitary equivalent states $\{\rho^{(0)}, \rho^{(1)},
\cdots \rho^{(\Delta)}\}$ (with $\rho^{(0)}=\rho$), such that $\rho^{(b)}=({\overline{U}^{(b,b^\prime)}}^\dagger\otimes {U^{(b,b^\prime)}}^\dagger) \rho^{(b^\prime)}(\overline{U}^{(b,b^\prime)}\otimes U^{(b,b^\prime)})$. For each state $\rho^{(b)}$, we then search for the optimum rotation matrix $\mathcal{O}^{(b)}$ such that the corresponding observable $\mathcal{M}^{(b)}$ has maximum average over $\rho^{(b)}$.

Note that for an arbitrary separable or entangled state, each term on the RHS of Eq. \eqref{RhoMkappa2} is bounded from above by $\kappa-1/d$, i.e. $\Tr\{ \rho^{(b)}\mathcal{M}^{(b)}(\kappa)\}\le \kappa-1/d$ for  arbitrary  state $\rho^{(b)}$ and $b=0,1,\cdots,\Delta$. This, in turn, implies that  no entanglement can be detected based on a single pair of measurements \cite{ChruscinskiPRA2018}. It turns out, however,  that  for a pair of measurements (say $b=0$), the bound can always  be achieved  by any  pure state, regardless of whether it is entangled or separable. To show this,  suppose $\ket{\psi}=\sum_{i=0}\sqrt{\lambda_i}\ket{e_ie_i}$ is  an arbitrary  pure state of bipartite $d\otimes d$ system in its Schmidt form. Suppose also that the traceless matrices $M_n^{(0)}$ are diagonal in the local Schmidt basis $\{\ket{e_i}\}_{i=0}^{d-1}$. Then
\begin{eqnarray}\label{Rho0M0}
\Tr\{ \rho^{(0)}\mathcal{M}^{(0)}(\kappa)\}&=&\sum_{i=0}^{d-1}\lambda_i\sum_{m,n=0}^{d-1}\mathcal{O}^{(0)}_{mn}\mu_{n\oplus i}\mu_{m\oplus i} \\ \nonumber
&=&\sum_{i=0}^{d-1}\lambda_i\sum_{n=0}^{d-1}\mu^2_{n\oplus i}=\kappa-1/d,
\end{eqnarray}
where the second equality follows by choosing the rotation matrix as $\mathcal{O}^{(0)}=\Id$, and the third equality comes from the normalization condition $\sum_{i=0}^{d-1}\lambda_i=1$ and Eq. \eqref{POVMconditions4} with $n=n^\prime$.  That  $\kappa-1/d$ is, indeed,  a bound  follows from the fact $\Tr\{ \rho^{(0)}\mathcal{M}^{(0)}(\kappa)\}\le \max\{\textrm{Spect}\{\mathcal{M}^{(0)}(\kappa)\}\}\le \max\{\textrm{Spect}\{[M_n^{(0)}(\kappa)]^2\}\}\le \Tr\{[M_n^{(0)}(\kappa)]^2\}=\kappa-1/d$, where  the first inequality can be saturated only by pure states.

 The arguments given above show that for an arbitrary pure state $\ket{\psi}$, one can always achieve the upper bound \eqref{Rho0M0}, just by choosing a pair of measurements diagonal in the local Schmidt basis. For any other pair of measurements complementary to the first one,  the corresponding contribution is nonzero if and only if the pure state is entangled. To show this, we suppose $M_n^{(0)}$ is diagonal in the local Schmidt basis and write $M_n^{(1)}=U^{(1)}M_n^{(0)}{U^{(1)}}^\dagger$, where $U^{(1)}$ is the discrete Fourier transformation defined by  $[U^{(1)}]_{kl}=\omega^{kl}$ with $\omega=\e^{2\pi i/d}$.  We obtain for $b=1$
 \begin{eqnarray}\label{Rho1M1-1}\nonumber
 \Tr\{\rho^{(1)}\mathcal{M}^{(1)}\}&=&\frac{1}{d^2}\sum_{m,n=0}^{d-1}\sum_{k,l=0}^{d-1}\sum_{r,s=0}^{d-1}\mathcal{O}^{(1)}_{mn}[M^{(0)}_n]_{rr}[M^{(0)}_m]_{ss} \;\;\\ \nonumber
 & & \qquad\qquad\qquad\quad \times \sqrt{\lambda_k\lambda_l}\omega^{(l-k)(r+s)} \\
  &=&\frac{1}{d^2}\sum_{m,n=0}^{d-1}\sum_{r,s=0}^{d-1}\mathcal{O}^{(1)}_{mn}[M^{(0)}_n]_{rr}[M^{(0)}_m]_{ss} \;\;\\ \nonumber
   & & \qquad\qquad \times \left|\sum_{j=0}^{d-1}\sqrt{\lambda_j}\e^{-2\pi i(r+s)j/d}\right|^2.
  \end{eqnarray}
 Using Eq. \eqref{Qmatrix}, one can write $[M^{(0)}_n]_{rr}=(\mathcal{Q}_{n,d-r}-1/d)/q$ and $[M^{(0)}_m]_{ss}=(\mathcal{Q}_{m,d-s}-1/d)/q$. Inserting these in the equation above, we obtain  after some calculation
\begin{eqnarray}\label{Rho1M1-2}
 \Tr\{\rho^{(1)}\mathcal{M}^{(1)}(\kappa)\}&=&\frac{1}{q^2 d^2}\sum_{r,s=0}^{d-1}\left(\tilde{\mathcal{O}}^{(1)}_{d-s,d-r}-1/d\right) \;\;\\ \nonumber  & & \qquad \times \left|\sum_{j=0}^{d-1}\sqrt{\lambda_j}\e^{-2\pi i(r+s)j/d}\right|^2,
  \end{eqnarray}
where we have defined the new orthogonal matrix $\tilde{\mathcal{O}}^{(1)}=\mathcal{Q}^\T\mathcal{O}^{(1)}\mathcal{Q}$ as an  orthogonal transformation of $\mathcal{O}^{(1)}$ by means of  $\mathcal{Q}$.  The square term  can be written as
$1+\sum_{j\ne k}\sqrt{\lambda_j\lambda_k}\cos\{2\pi(r+s)(k-j)/d\}$. It turns out that $\cos\{2\pi(r+s)(k-j)/d\}$ takes its maximum value 1 whenever $r+s=0$ or $r+s=d$. Using this in Eq.  \eqref{Rho1M1-2}, we find
\begin{eqnarray}\label{Rho1M1-3}
\Tr\{\rho^{(1)}\mathcal{M}^{(1)}(\kappa)\}&=&\frac{1}{q^2 d^2}\sum_{r=0}^{d-1}\tilde{\mathcal{O}}^{(1)}_{r,d-r}\sum_{j\ne k}\sqrt{\lambda_j\lambda_k} \\ \nonumber
&+&\frac{1}{q^2 d^2}\sum_{s=0}^{d-1}\sum_{r\ne d-s}^{d-1}\tilde{\mathcal{O}}^{(1)}_{d-s,d-r}\sum_{j\ne k}\sqrt{\lambda_j\lambda_k} \\ \nonumber & &\quad \qquad \times\; \cos\{2\pi(r+s)(k-j)/d\}.
\end{eqnarray}
By choosing the nonzero entries of the orthogonal matrix $\tilde{\mathcal{O}}^{(1)}$ as  $\tilde{\mathcal{O}}^{(1)}_{r,d-r}=1$ for $r=0,\cdots,d-1$, the first summation takes its maximum value and the second one vanishes. A similar result can be obtained for all other complementary measurements $b\ne 0$.
  We therefore provide the following proposition.
\begin{proposition}\label{Prop2}
For an arbitrary pure bipartite $d\otimes d$ state $\ket{\psi}$ with Schmidt decomposition $\ket{\psi}=\sum_{i=0}\sqrt{\lambda_i}\ket{e_ie_i}$, we have
\begin{eqnarray}\label{RhobMb}
\Tr\{ \rho^{(0)}\mathcal{M}^{(0)}(\kappa)\}&=&(\kappa-1/d), \\
\Tr\{ \rho^{(b)}\mathcal{M}^{(b)}(\kappa)\}&=&(\kappa-1/d)E(\psi),
\end{eqnarray}
for $b=1,\cdots,\Delta$, where $E(\psi)=\sum_{j\ne k}\sqrt{\lambda_j\lambda_k}/(d-1)$ is an entanglement monotone, ranging from zero for product states to its maximum value $1$ for maximally entangled states.
\end{proposition}
It remains only to prove that $E(\psi)$ is an entanglement monotone, i.e. it is a Schur concave function of the Schmidt numbers $\lambda_i$s, meaning that it is invariant under permutation of the Schmidt numbers and    $(\lambda_i-\lambda_j)\left(\partial E/\partial \lambda_i-\partial E/\partial \lambda_j\right)\le 0$ for all pairs $i,j$. In our case one can easily  find that $(\lambda_i-\lambda_j)\left(\sum_{k\ne i}\sqrt{\lambda_k}/\sqrt{ \lambda_i}-\sum_{k\ne j}\sqrt{\lambda_k}/\sqrt{ \lambda_j}\right)\le 0$, so  $E(\psi)$ is an entanglement monotone. According to this result,  to detect entanglement of a general pure state, we need only two pairs of MUMs. This, in some sense, is  a generalization of the result provided by condition \eqref{Wit1} in \cite{SpenglerPRA2012} based on the MUBs.
We now provide  some examples.

{\it Example 1.---}We first consider the isotropic states. An   isotropic state of a $d\otimes d$ system is invariant under any unitary transformation of the form $\overline{U}\otimes U$, and can be written as
\begin{equation}
\rho_{I}=\alpha\left|\phi_{d}^{+}\right\rangle\left\langle\phi_{d}^{+}\right|+\frac{1-\alpha}{d^{2}} \Id_{d}\otimes \Id_{d},
\end{equation}
where $ \ket{\phi_{d}^{+}}=\dfrac{1}{\sqrt{d}}\sum_{i=0}^{d-1}\ket{ii}$ and $0 \leq \alpha<1$. It is known that isotropic states are entangled if and only if $\alpha>1 /(d+1)$ \cite{BertlmannPRA2005}.  Using condition \eqref{Wit1},  entanglement detection of isotropic states is studied in \cite{SpenglerPRA2012}. In \cite{LiITP2019}, the authors have used a complete set of MUMs  provided in \cite{KalevNJP2014}, and  performed a complete detection of entanglement of these states using \eqref{WitnessP}. Using Eq. \eqref{RhoMkappa2},  we find
\begin{eqnarray}
\Tr\{\rho_I \mathcal{M}(\kappa)\}&=&\alpha \sum_{b=0}^{{\Delta}}\sum_{k,l=0}^{d-1}\mathcal{O}_{kl}^{(b)}\bra{\phi_{d}^{+}}\overline{M}_{l}^{(0)}\otimes M_{k}^{(0)}\ket{\phi_{d}^{+}} \;\;\;\\ \nonumber
&=&\frac{\alpha}{d} \sum_{b=0}^{{\Delta}}\sum_{k,l=0}^{d-1}\mathcal{O}_{kl}^{(b)}\Tr\{M_{l}^{(0)} M_{k}^{(0)}\}
\\ \nonumber
&=&\alpha\left[\frac{\kappa -1/d}{d-1}\right] \sum_{b=0}^{{\Delta}}\left[\Tr\{\mathcal{O}^{(b)}\}-1\right],
\end{eqnarray}
where in the last line we have used Eq. \eqref{TrMMp-1} and that $\sum_{k,l=0}^{d-1}\mathcal{O}_{kl}^{(b)}=d$. Obviously, the optimum detection happens when $\Tr\{\mathcal{O}^{(b)}\}$ takes its maximum value $d$, i.e. when $\mathcal{O}^{(b)}=\Id_d$ for all $b$.   Using these  in Eq. \eqref{Prop1}, we get $\Tr\{\rho_I W(\kappa)\}=(\kappa -1/d)(1-\alpha({\Delta}+1))$, implying that the state is entangled when $\alpha>1/({\Delta}+1)$. This result can be  obtained also if one uses the traceless property of $\mathcal{M}(\kappa)$ together with the results provided by Proposition \ref{Prop2}.  It follows that the complete detection  of entanglement of the isotropic states occurs when ${\Delta}=d$, i.e. there exists a complete set of $d+1$ MUMs \cite{SpenglerPRA2012,LiITP2019}.

{\it Example 2.---}For the second example, we consider a noisy Dicke state  defined by
\begin{equation}\label{NoisyDickeS}
\rho=(1-p) \ket{D_{N}^{k}}\bra{D_{N}^{k}}+ p\dfrac{\Id_{2^N}}{2^N},
\end{equation}
where $\ket{D_{N}^{k}}$ is an $N$-qubit Dicke state with $k$ excitation
\begin{eqnarray}\label{DickeS-1}
\ket{D_{N}^{k}}= \binom{N}{k}^{-1/2} \sum_{l}  \mathcal{P}_{l} \left\lbrace  \ket{0}^{\otimes N-k} \otimes \ket{1}^{\otimes k} \right\rbrace.
\end{eqnarray}
Above,  $\sum_{l} \mathcal{P}_{l}\{\cdot\}$ denotes the sum over all possible permutations. In what follows we suppose that the number of qubits is even, i.e. $N=2n$ for some integer $n$. Accordingly, for a balanced bipartition $(n|n)$, a Dicke state can be expressed in Schmidt form as \cite{MorenoARXIV2018}
\begin{eqnarray}\label{DickeS-2}
\ket{D_{N=2n}^{k}}= \sum_{q=q^\prime}^{q^{\prime\prime}}\sqrt{\lambda_q}\ket{D_{n}^{q}}\ket{D_{n}^{q-k}},
\end{eqnarray}
where  the  Schmidt coefficients are given by
\begin{eqnarray}\label{DickeS-Shcmidt}
\lambda_{q}=\frac{N!}{\binom{N}{k}\binom{N}{n}}\frac{1}{q!(n-q)!(k-q)!(n-k+q)!},
\end{eqnarray}
and $q^{\prime}=\max\{0,k-n\}$,   $q^{\prime\prime}=\min\{n,k\}$, and $q^\prime<q^{\prime\prime}$.
Since the dimension $d=2^n$ is a power of prime number, there exists a complete set of $d+1$ MUMs, i.e. $\Delta=d$. In this case, the witness detects the state as entangled if
\begin{eqnarray}
p<\frac{E(D_{2n}^{k})}{E(D_{2n}^{k})+2^{-n}}.
\end{eqnarray}
For a four-qubit Dicke state with two excitation, the entanglement reads $E(D_{4}^{2})=5/9$, so the state is entangled up to the noise threshold of $p<20/29$.

{\it Example 3.---}As another example, we consider the following PPT entangled  state introduced  in \cite{ChruscinskiPRA2018}
\begin{eqnarray}\label{H3x3}
\rho=\dfrac{1}{15}
\begin{pmatrix}
\begin{array}{c c c|c c c | c c c}
1&0&0&0&1&0&0&0&1\\
0&2&0&0&0&-1&-1&0&0\\
0&0&2&-1&0&0&0&-1&0\\ \hline
0&0&-1&2&0&0&0&-1&0\\
1&0&0&0&1&0&0&0&1\\
0&-1&0&0&0&2&-1&0&0\\ \hline
0&-1&0&0&0&-1&2&0&0\\
0&0&-1&-1&0&0&0&2&0\\
1&0&0&0&1&0&0&0&1
\end{array}
\end{pmatrix}.
\end{eqnarray}
A simple calculation shows that $\rho$ has 5 nonzero eigenvalues, all equal to $1/5$.
The use of a complete set of four MUBs ($\kappa=1$) to detect entanglement of this state has been studied in \cite{ChruscinskiPRA2018}, however, as we show below it is always possible to detect entanglement of $\rho$ by choosing a suitable set of three MUMs (or MUBs).
In this case we find
\begin{eqnarray}\nonumber
\Tr\{\rho \mathcal{M}(\kappa)\}=\frac{1}{10}(\kappa-1/3)\left\{\left[1-\Tr{\mathcal{O}^{(0)}}\right]+\left[1-\Tr{\mathcal{O}^{(1)}}\right]\right.  \\ \nonumber
-\left.2\left[1-\Tr{\mathcal{O}^{(2)}}\right]-2\left[1-\Tr{\mathcal{O}^{(3)}}\right]
\right\},
\end{eqnarray}
where brackets on the RHS correspond to terms  $b=0,1,2,3,$ of Eq. \eqref{RhoMkappa2}, respectively.
Using this  in Eq. \eqref{Prop1}, we get
\begin{eqnarray}
\Tr\{\rho W(\kappa)\}=\frac{1}{10}(\kappa-1/3)\left\{12+\left[\Tr{\mathcal{O}^{(0)}}+\Tr{\mathcal{O}^{(1)}}\right]\right.\;\;\;\; \\ \nonumber
-\left.2\left[\Tr{\mathcal{O}^{(2)}}+\Tr{\mathcal{O}^{(3)}}\right]
\right\}.
\end{eqnarray}
To obtain the optimum detection, we have to know $\Tr\{\mathcal{O}\}$ for a general $3\times 3$ orthogonal matrix  $\mathcal{O}$.   Given a unit vector $\hat{\boldsymbol{n}}=(n_{1},n_{2},n_{3})$ and an angle $\theta$, a general rotation matrix in $\mathbb{R}^3$ is given by $R(\hat{\boldsymbol{n}},\theta)=[R_{ij}]$  where
$R_{ij}=n_in_j(1-\cos{\theta})+\delta_{ij}\cos{\theta}-\epsilon_{ijk}n_k\sin{\theta}$, and $\epsilon_{ijk}$ is the levi-civita symbol and sum over the repeated index is understood. Obviously, rotations  preserve the rotation axis, and in particular we are interested in the case where the rotation preserves the unit vector  $\hat{\boldsymbol{n}}_{\ast}=\dfrac{1}{\sqrt{3}}(1,1,1)$. In this case we obtain $\Tr\{\mathcal{O}(\theta)\}=1+2\cos{\theta}$.
Clearly, the optimum detection happens when $\Tr\{\mathcal{O}^{(0)}\}=\Tr\{\mathcal{O}^{(1)}\}=-1$ and $\Tr\{\mathcal{O}^{(2)}\}=\Tr\{\mathcal{O}^{(3)}\}=2$, i.e. when $\theta_{0}=\theta_{1}=\pi $ and $ \theta_{2}=\theta_{3}=0$. Accordingly, in order to construct a  witness to detect entanglement of  the state \eqref{H3x3},  we need only a set of three MUMs $\{\mathcal{P}^{(b)}\}$ for $b=0,2,3$ or $b=1,2,3$, for which $\Tr\{\rho W(\kappa)\}=-\frac{1}{10}(\kappa-1/3)$.   For a fixed  purity $\kappa$, using a complete set of four MUMs, however, leads to  more negative value $-\frac{1}{5}(\kappa-1/3)$.   Although the full detection of entanglement occurs for  any purity $\kappa>1/3$,  the depth of detection is increased by increasing $\kappa$.

{\it Example 4.---}As  a final example, we consider first a rank-two  bipartite state described by the following ensemble
\begin{eqnarray}\label{Example4-1}
\rho=p_0\ket{\psi^{(0)}}\bra{\psi^{(0)}}+p_1\ket{\psi^{(1)}}\bra{\psi^{(1)}},
\end{eqnarray}
where the  pure states $\ket{\psi^{(0)}}$ and $\ket{\psi^{(1)}}$  have the Schmidt decomposition $\ket{\psi^{(0)}}=\sum_{n=0}\sqrt{\lambda^{(0)}_n}\ket{e^{0}_n e^{0}_n}$ and $\ket{\psi^{(1)}}=\sum_{n=0}\sqrt{\lambda^{(1)}_n}\ket{e^{1}_n e^{1}_n}$, respectively,  such that   $|\braket{e^{0}_n}{ e^{1}_{n^\prime}}|=1/\sqrt{d}$.   If we use only two pairs of MUMs corresponding to the two Schmidt basis, we obtain from the results of the  Proposition \ref{Prop2}
\begin{eqnarray}
\Tr\{\rho \mathcal{M}(\kappa)\}=(\kappa-1/d)\left(1+p_0E(\psi^{(0)})+p_1E(\psi^{(1)})\right).
\end{eqnarray}
It turns out that  the state is entangled if and only if at least one of the pure states  $\ket{\psi^{(0)}}$ and $\ket{\psi^{(1)}}$ is entangled. Moreover, for such mixtures, two pairs of MUMs are enough to decide whether  the state is entangled or not.  Any attempt to increase the number of MUMs leads to a corresponding increase in the contribution of  the second term, without having any effect on the detection of the entanglement of the state.  We can extend this result by considering the following  mixture
\begin{eqnarray}\label{Example4-2}
\rho=\sum_{b=0}^{\Delta^\prime}p_b\ket{\psi^{(b)}}\bra{\psi^{(b)}},
\end{eqnarray}
where  the local Schmidt basis of   $\{\ket{\psi^{(b)}}\}$ are mutually unbiased, i.e.  for $\ket{\psi^{(b)}}=\sum_{n=0}\sqrt{\lambda^{(b)}_n}\ket{e^{b}_n e^{b}_n}$ we have  $|\braket{e^{b}_n}{ e^{b^\prime}_{n^\prime}}|=1/\sqrt{d}$ for $b\ne b^\prime$.
Using the results of the Proposition \ref{Prop2}, we  find
\begin{eqnarray}
\Tr\{ \rho^{(b)}\mathcal{M}^{(b)}(\kappa)\}=(\kappa-1/d)\left[p_b+\sum_{i\ne b}^{\Delta^\prime}p_iE(\psi^{(i)})\right],
\end{eqnarray}
for $b=0,1,\cdots,\Delta^\prime$ and
\begin{eqnarray}
\Tr\{ \rho^{(b)}\mathcal{M}^{(b)}(\kappa)\}=(\kappa-1/d)\sum_{i=0}^{\Delta^\prime}p_iE(\psi^{(i)}),
\end{eqnarray}
for $b=\Delta^\prime+1,\cdots,\Delta$. Accordingly, with the assumption that $\Delta^\prime \le \Delta$, we get
\begin{eqnarray}
\Tr\{\rho \mathcal{M}(\kappa)\}=(\kappa-1/d)\left[1+\Delta\sum_{b=0}^{\Delta^\prime}p_bE(\psi^{(b)})\right].
\end{eqnarray}
 We find again that  the  ensemble \eqref{Example4-2} is entangled if and only if one of the pure states $\ket{\psi^{(b)}}$ is entangled.  However, for such mixtures, depending  on the fraction and the  entanglement of the pure states, the number of MUMs needed to detect entanglement of the mixture is different ranging from two to $\Delta^\prime+1$. For example, when all the pure states are maximally entangled state, i.e. $E(\psi^b)=1$ for $b=0,\cdots,\Delta^\prime$, two pairs of MUMs suffice to  detect entanglement of the mixture. On the other hand, for a mixture of one maximally entangled state and $\Delta^\prime$ disentangled states, say $E(\psi^0)=1$ and $E(\psi^{b})=0$ for $b=1\cdots,\Delta^\prime$, two pairs of MUMs is enough  only if $2p_0+p_1>1$.

\section{Conclusion}\label{SectionConclusion}

 We have presented a method to construct   a class of MUMs that encompasses the full range of purity, from $1/d$ for totally mixed measurements  to $1$ for MUBs.  Mutual compatibility of different outcomes within each measurement and the mutually unbiasedness of different measurements are two important common features of our class  and MUBs.
 This similarity prevents one to construct a complete set of these MUMs. In contrary to MUBs, however,  the spectra of such  MUMs are nontrivial, and can be expressed in terms of $\left\lfloor \frac{d-1}{2} \right\rfloor$ independent parameters. This spectra provides a way to construct a class of orthogonal matrices with the property that they leave the vector  $\boldsymbol{n}_\ast=(1,1,\cdots,1)/\sqrt{d}$ invariant. For a fixed purity $\kappa$, these orthogonal matrices depend on $\left\lfloor \frac{d-1}{2} \right\rfloor$ independent parameters.  We used this class of MUMs to study entanglement detection of $d\otimes d$ bipartite states. We show that to detect entanglement of a general bipartite pure state, two MUMs are suffice; The first one assigns  a zero mean value for all pure states, however, a complementary measurement is needed to give a negative mean value  for entangled states. We  show that the function described  this negative value is Schur concave of Schmidt numbers so that can be regarded as  an entanglement monotone.  For bipartite mixed states, however,   the  number of MUMs which is needed to detect entanglement increases in general. We examine this by providing some examples and show that although for a full rank state, such as isotropic state, a complete set of MUMs is needed to completely detect entanglement, for mixed states with low-dimensional support a smaller set of MUMs suffices to capture entanglement.

\section*{acknowledgment}
The authors would like to thank Fereshte Shahbeigi  for helpful discussion and comments. This work was supported by Ferdowsi University of Mashhad under Grant No.  3/47501 (1397/6/27).

\appendix
\section{The number of independent eigenvalues}\label{App1}
To count the number of independent eigenvalues, we have to find the number of independent constraints on the set of eigenvalues. Clearly, Eq. \eqref{POVMconditions3} gives one independent relation.  Equation \eqref{POVMconditions4}, on the other hand,  reads
\begin{equation} \label{EigenCondition1}
\sum_{j=0}^{d-1} \mu^2_{n\oplus j}=\kappa-1/d,
\end{equation}
for $n=n^\prime$, and
\begin{equation}\label{EigenCondition2}
\sum_{j=0}^{d-1} \mu_{n\oplus j}\mu_{n^\prime\oplus j}=-\left(\frac{\kappa -1/d}{d-1}\right),
\end{equation}
for $n\ne n^\prime$. Equation \eqref{EigenCondition1} gives its own independent relation, however,  Eq. \eqref{EigenCondition2} provides  independent relations  only for $n^\prime=n\oplus 1,n\oplus 2,\cdots,n\oplus \left\lfloor d/2\right\rfloor$, where $\left\lfloor d/2\right\rfloor$ denotes the integral part of $d/2$, i.e.  $\left\lfloor d/2\right\rfloor=d/2$ if $d$ is even and  $\left\lfloor d/2\right\rfloor=(d-1)/2$ if $d$ is odd.  This follows from the fact the LHS of  this equation is invariant under the change $\{n \rightarrow n+n_0,\; n^\prime \rightarrow n^\prime+n_0\}$ for any  $n_0=0,\cdots d-1$.  For $n^\prime=1,\cdots,\left\lfloor(d-1)/2\right\rfloor$, the multiplicity is $d$, however, for $n^\prime=d/2$ (when $d$ is even) the corresponding multiplicity is $d/2$.  There is, however, a  relation between Eqs. \eqref{EigenCondition1} and \eqref{EigenCondition2}, follows easily from
\begin{equation}
2\sum_{n<n^\prime}^{d-1} \mu_{n}\mu_{n^\prime}=\left(\sum_{n=0}^{d-1} \mu_{n}\right)^2-\sum_{n=0}^{d-1} \mu^2_{n}=-\left(\kappa -1/d\right),
\end{equation}
where, using the multiplicity of each term,  can be written  as
\begin{eqnarray}
d\sum_{i=1}^{\left\lfloor\frac{d-1}{2}\right\rfloor}\left[\sum_{j=0}^{d-1} \mu_{n\oplus j}\mu_{n\oplus i+ j}\right]+\frac{d}{2}\mathcal{E}=-\frac{1}{2}\left(\kappa d-1\right).
\end{eqnarray}
Above $\mathcal{E}=\sum_{j=0}^{d-1} \mu_{n\oplus j}\mu_{n\oplus d/2+ j}$ if $d$ is even and it is zero if $d$ is odd.
Putting everything together, we find  the number of independent eigenvalues as $N=d-(2+\left\lfloor d/2\right\rfloor-1)=\left\lfloor\frac{d-1}{2}\right\rfloor$.


\begin{thebibliography}{99}


\bibitem{SchwingerPNAS1960} Schwinger, J.: Unitary operator bases. Proc. Natl. Acad. Sci. USA \textbf{46},  570 (1960)

\bibitem{WoottersAP1989}  Wootters, W.K., Fields, B.D.: Optimal state-determination by mutually unbiased measurements. Ann. Phys. (NY) \textbf{191}, 363 (1989)

\bibitem{IvanovicJPA1981}  Ivanovi{\'{c}}, I.D.: Geometrical description of quantal state determination. J. Phys. A: Math. Gen.  \textbf{14}, 3241 (1981)

\bibitem{KlappeneckerPtoc2005}  Klappenecker, A., R\"{o}tteler, M.: Mutually unbiased bases are complex projective 2-designs. Proc. Int. Symp. Inf. Theory.   pp 1740–4 (2005)

\bibitem{GottesmanPRA1996}
 Gottesman, D.: Class of quantum error-correcting codes saturating the quantum Hamming bound. Phys. Rev. A. \textbf{54}, 1862 (1996)

\bibitem{CalderbankPRL1997}
 Calderbank, A.R., Rains, E.M., Shor, P.W., Sloane, N.J.A.: Quantum error correction and orthogonal geometry. Phys. Rev. Lett. \textbf{78}, 405 (1997)

\bibitem{BrusPRL2002}
 Bru{\ss}, D., Macchiavello, C.: Optimal eavesdropping in cryptography with three-dimensional quantum states. Phys. Rev. Lett. \textbf{88}, 127901 (2002)

\bibitem{CerfPRL2002}
 Cerf, N.J., Bourennane, M., Karlsson, A., Gisin, N.: Security of quantum key distribution using d-level systems. Phys. Rev. Lett. \textbf{88}, 127902 (2002)

\bibitem{AdamsonPRL2010}
Adamson, R.B.A., Steinberg, A.M.: Improving quantum state estimation with mutually unbiased bases. Phys. Rev. Lett. \textbf{105}, 030406 (2010)

\bibitem{SpenglerPRA2012}
 Spengler, C., Huber, M., Brierley, S., Adaktylos, T., Hiesmayr, B.C.: Entanglement detection via mutually unbiased bases. Phys. Rev. A. \textbf{86}, 022311 (2012)

\bibitem{EnglertPLA2001}
Englert, B-G., Aharonov, Y.: The mean king's problem: prime degrees of freedom. Phys. Lett. A. \textbf{284}, 1 (2001)

\bibitem{AravindZN2003}
 Aravind, P.K.: Solution to the king's problem in prime power dimensions. Z. Naturforsch.  \textbf{58a}, 85 (2003)

\bibitem{ButterleyPLA2007}
 Butterley, P., Hall, W.: Numerical evidence for the maximum number of mutually unbiased bases in dimension six. Phys. Lett. A. \textbf{369}, 5 (2007)

 \bibitem{DurtIJQI2010}
  Durt, T., Englert, B-G, Bengtsson, I., \.{Z}yczkowski, K.: On mutually unbiased bases. Int. J Quantum Inf. \textbf{8}, 535 (2010)

\bibitem{KalevNJP2014}
 Kalev, A., Gour, G.: Mutually unbiased measurements in finite dimensions. New J. Phys. \textbf{16}, 053038 (2014)

\bibitem{GraydonJPA2016-1}
 Graydon, M.A., Appleby, D.M.: Quantum conical designs. J. Phys. A: Math. Theor. \textbf{49}, 085301 (2016)

\bibitem{BransdenPRA2016}
 Bransden, S., Arno, M.D., Szymusiak, A.: Communication capacity of mixed quantum t-designs. Phys. Rev. A. \textbf{94}, 022335 (2016)


\bibitem{ZyczkowskiBook2017}
 Bengtsson, I., \.{Z}yczkowski, K.: Geometry of Quantum States: An Introdunction to Quantum Entanglement.  Cambridge University Press, Cambridge (2017)

\bibitem{ChenPRA2014}
    Chen, B., Ma, T., Fei, S-M.: Mutually unbiased measurement based entanglement witnesses. Phys. Rev. A. \textbf{6}, 064302 (2014)

 \bibitem{ChruscinskiPRA2018}
    Chru\'{s}ci\'{n}ski, D., Sarbicki, G., Wudarski, F.: Entanglement witnesses from mutually unbiased bases. Phys. Rev. A. \textbf{97}, 032318 (2018)

   \bibitem{LiITP2019}
   Li, T., Lai, L., Fei, S., Wang, Z.: Mutually unbiased measurement based entanglement witnesses. Int. J. Theor. Phys. \textbf{58}, 3973{-}3985 (2019)

   \bibitem{BertlmannPRA2005}
 Bertlmann, R.A., Durstberger, K., Hiesmayr, B.C., Krammer, P.: Optimal entanglement witnesses for qubits and qutrits. Phys. Rev. A. \textbf{72}, 052331 (2005)

\bibitem{MorenoARXIV2018}
 Moreno, M.G.M., Fernando Parisio: All bipartitions of arbitrary Dicke states. arXiv: 1801.00762 (2018)




 \end{thebibliography}
\end{document}